\documentclass{emulateapj}
\usepackage{apjfonts}
\usepackage{graphicx}

\usepackage{amsmath}

\shorttitle{Formation of Massive Molecular Cloud Cores}
\shortauthors{T. INOUE and Y. FUKUI}

\begin{document}

\title{
FORMATION OF MASSIVE MOLECULAR CLOUD CORES BY CLOUD-CLOUD COLLISION
}
\author{Tsuyoshi Inoue\altaffilmark{1} and Yasuo Fukui\altaffilmark{2}}
\altaffiltext{1}{Department of Physics and Mathematics, Aoyama Gakuin University, Sagamihara, Kanagawa 252-5258, Japan; inouety@phys.aoyama.ac.jp}
\altaffiltext{2}{Department of Physics, Graduate School of Science, Nagoya University, Furo-cho, Chikusa-ku, Nagoya 464-8602, Japan}

\begin{abstract}
Recent observations of molecular clouds around rich massive star clusters including NGC3603, Westerlund 2, and M20 revealed that the formation of massive stars could be triggered by a cloud-cloud collision.
By using three-dimensional, isothermal, magnetohydrodynamics simulations with the effect of self-gravity, we demonstrate that massive, gravitationally unstable, molecular cloud cores are formed behind the strong shock waves induced by the cloud-cloud collision.
We find that the massive molecular cloud cores have large effective Jeans mass owing to the enhancement of the magnetic field strength by shock compression and turbulence in the compressed layer.
Our results predict that massive molecular cloud cores formed by the cloud-cloud collision are filamentary and threaded by magnetic fields perpendicular to the filament.
\end{abstract}


\section{Introduction}
Massive stars are astronomically very interesting objects that affect the surrounding medium by strong radiation and stellar wind, cause supernovae and probably gamma-ray bursts at the end of their lives, and leave behind neutron stars and black holes.
However, how massive stars are formed in molecular clouds is a matter of debate (e.g., Zinnecker \& Yorke 2007).
Theoretical studies suggest that massive stars can be formed by gravitational collapse of massive molecular cloud cores (Nakano et al. 2000; McKee \& Tan 2002; Yorke \& Sonnhalter 2002; Krumholz et al. 2009) and/or by competitive gas accretion in the stellar cluster-forming region (Bonnell et al. 2004).
Such theoretical studies mostly focused on massive star formation in a given gravitationally unstable core.
Thus, to draw the complete scenario of massive star formation, we need to clarify how massive gravitationally unstable objects are formed in a molecular cloud.

CO observations by using the NANTEN2 telescope have revealed two distinct molecular clouds in the star burst cluster NGC3603, which includes more than 30 O stars, and it has been shown that the cloud-cloud collision trigger the cluster formation in the shock-compressed layer (Fukui et al. 2013, submitted to ApJ).
The NANTEN2 observations revealed two more cases of massive star formation by similar cloud-cloud collisions in Westerlund 2 (Furukawa et al, 2009; Ohama et al. 2010) and in M20 (Torii et al. 2011), suggesting the importance of such triggering mechanisms in massive star formation.
One of the most outstanding features of massive star-forming cloud-cloud collisions is the typically large relative speed of the collisions (20 km s$^{-1}\sim 100$ in sonic Mach number), while the colliding clouds have a variety of mass and size.

The cloud-cloud collision was studied using hydrodynamics simulations by Habe \& Ohta (1992), Anathpindika (2010) and Duarte et al. (2011) (see also, Miniati et al. 1999 for collision of HI clouds).
Habe \& Ohta (1992) and Anathpindika (2010) found that the cloud-cloud collision induces the formation of cloud cores by enhanced self-gravity as a consequence of shock compression.
In previous studies, the role of the magnetic field on core formation was neglected.
If we omit the magnetic field, the shock compression due to the strong collision tends to generate cores with lower mass, because the density enhancement generally leads to smaller Jeans mass.
However, if we consider the effect of the magnetic field, the mass of the cores formed behind the shock can be enhanced owing to the strengthened magnetic field and enhanced effective Jeans mass.
In this study, by using three-dimensional, isothermal, magnetohydrodynamics (MHD) simulations with the effect of self-gravity, we examine the cloud-cloud collision and the formation of gravitationally unstable objects in the shock compressed layer.
The above mentioned observations of massive star formation sites and the discussion in Duarte et al. (2011) regarding the morphology of clouds formed by the cloud-cloud collision agree with the results of previous simulations.
Thus, in this letter, we focus on the shocked interface induced by the cloud-cloud collision and concentrate on how massive cores are formed in the shocked layer.

\section{Setup of Simulations}
We solve the ideal MHD equations with self-gravity, and assume an isothermal equation of state with a sound speed of $c_{\rm s}=0.2$ km s$^{-1}$, which corresponds to a medium of temperature $T=12$ K and mean molar weight of $2.4$.
The basic MHD equations are solved by using the second-order Godunov-type finite-volume scheme (van Leer 1979) that implements the approximate MHD Riemann solver developed by Sano et al. (1999).
The consistent method of characteristics with the constrained transport technique is used for solving the induction equations (Clarke 1996).
The integration of the ideal MHD part of the basic equations is performed in the conservative fashion.
The Poisson equation for self-gravity is solved by using the multi-grid method (William et al. 1988), in which the boundary condition is computed by the method developed by Miyama et al. (1987).

\begin{deluxetable*}{llllllll} \label{t1}[t]
%\tablewidth{0pt}
\tablecaption{Model Parameters}
\tablehead{Model No. & 1 & 2 & 3 & 4 & 5 & 6 & 7 }
\startdata
$\langle n\rangle_0$ [cm$^{-3}$] & 300 & 300 & 300 & 1000 & 300 & 300 & 300  \\
$\Delta n/\langle n\rangle_0$ & 0.33 & 0.33 & 0.33 & 0.33 & 0.33 & 0.33 & 0.50 \\
$B_0$ [$\mu$G] & 20 & 20 & 20 & 20 & 10 & 20 & 20 \\
$v_{\rm coll}$ [km/s]  & 10 & 10 & 10 & 10 & 10 & 5.0 & 10  \\
Gravity & Yes & Yes & No & Yes & Yes & Yes & Yes \\
Resolution [pc] & 8.0/1024 & 8.0/512 & 8.0/1024 & 8.0/512 & 8.0/512 & 8.0/512 & 8/512  \\
$M_{\rm core,\,tot}$ [$M_{\rm sun}$] & 194 & 167 & N/A & 126 & 35 & 24 & 69\\
$M_{\rm mag}$ [$M_{\rm sun}$] & 122 & 124 & N/A & 63 & 14 & 36 & 34\\
$M_{\rm turb}$ [$M_{\rm sun}$] & 71 & 41 & N/A & 63 & 21 & 53 & 34\\
$\langle n\rangle_{\rm core}$ [cm$^{-3}$] & 8.4e4 & 8.3e4 & N/A & 4.9e5 & 3.1e5 & 2.0e4 & 1.7e5 \\
$\langle B\rangle_{\rm core}$ [$\mu$G] & 2.8e2 & 2.8e2 & N/A & 6.3e2 & 2.9e2 & 1.8e2 & 2.9e2\\
$\Delta v_{\rm core}$ [km s$^{-1}$] & 1.2 & 0.97 & N/A & 1.3 & 0.86 & 0.23 & 1.0 \\
$t_{\rm form}$ [ Myr ] & 0.63 & 0.63 & N/A & 0.29 & 0.36 & 0.78 & 0.39
\enddata
\end{deluxetable*}

We prepare a cubic numerical domain of side lengths $L_{\rm box}=8.0$ pc that consists of $1024^3$ or $512^3$ uniform numerical cells depending on the model (Table 1), indicating that the physical resolution is $\Delta x\simeq$  0.008 pc or 0.016 pc.
We initially set the isothermal gas with density fluctuations because molecular clouds are characterized by supersonic turbulence that induces density fluctuations (MacLow \& Klessen 2004; McKee \& Ostriker 2007; Ballesteros-Paredes et al. 2007).
The fluctuations are given as a superposition of sinusoidal functions with various wave numbers from $2\pi/L_{\rm box}\le|k|\le80\pi/L_{\rm box}$ and random phases (e.g., Inoue \& Inutsuka 2008 for the implementation).
The power spectrum of the density fluctuations is set as an isotropic power law $(\log\rho)_{k}^2\,k^2\propto k^{-2}$ (Beresnyak et al. 2005, see also, Elmegreen \& Scalo 2004; Scalo \& Elmegreen 2004), which is expected by supersonic turbulence that follows Larson's law (Larson 1981; Heyer \& Brunt 2004).
Thus, our initial density structure is parameterized by mean density $\langle n \rangle_0$ and dispersion $\Delta n\equiv (\langle n^2 \rangle-\langle n \rangle_0^2)^{1/2}$.
The model parameters are summarized in Table 1.
The random phases of the fluctuations are the same for all models to study the effects of the parameters for the same seed fluctuations.

To study the shocked layer generated by the cloud-cloud collision, we set the initial velocity $v_x(x)=v_{\rm coll}\tanh [(5-10\,x)/L_{\rm box}]$, i.e., two flows collide head on in the $y$-$z$ plane at the center of the numerical domain with relative velocity of $2\,v_{\rm coll}$.
At the $y$-$z$ boundary planes, the density is given by $\rho_{0}(x=v_{\rm coll}\,t,y,z)$ for the $x=8$ pc plane and  $\rho_{0}(x=L_{\rm box}-v_{\rm coll}\,t,y,z)$ for the $x=0$ pc plane, where $\rho_{0}(x,y,z)$ is the initial density field including fluctuations.
For the $x$-$y$ and $x$-$z$ boundary planes, periodic boundary conditions are imposed.
Because the shock waves induced by the collision are very strong, the resulting shocked layer is magnetized almost parallel to the shocked layer.
In addition, if collisions were random, the colliding clouds would have different orientations of magnetic fields.
Thus, the initial magnetic field is set as $\vec{B}=(0,B_0\,\cos[\pi/4],B_0\,\sin[\pi/4])$ for $x\le L_{\rm box}/2$ and $(0,B_0\,\cos[\pi/4],B_0\,\sin[-\pi/4])$ for $x>L_{\rm box}/2$.

It is widely known that artificial gravitational fragmentation can be avoided if the ratio of Jeans length to the numerical resolution is larger than four (Trulove et al. 1997).
Subsequently, we show that the effective Jeans length of the core is approximately 0.1 pc with an approximate resolution of 13 cell lengths for Model 1 and 6.5 cell lengths for the other models.

\section{Results}
\subsection{Fiducial model}
The colliding streams induce two shock waves at $x= L_{\rm box}/2$ plane that propagate oppositely.
In Model 1 (henceforth fiducial model), the sonic Mach number of the shock is approximately given by ${\cal M}_{\rm s}\simeq v_{\rm coll}/c_{\rm s}=50$ and the Alfv\'enic Mach number is ${\cal M}_{\rm A}\simeq v_{\rm coll}/\langle c_{\rm A}\rangle=12.3$.
According to the shock jump condition for the isothermal MHD shock, the compression ratio is given by
\begin{equation}\label{RH}
r=\rho_{\rm 1}/\rho_{\rm 0}=B_{\rm 1}/B_{\rm 0} \simeq \sqrt{2}\,{\cal M}_{\rm A},
\end{equation}
where subscript 0 (1) denotes the preshock (postshock) state.
Note that if we omit the effect of the magnetic field, the postshock density becomes $r \simeq {\cal M}_{\rm s}^2$ and is substantially different from the appropriate case of Eq.~(\ref{RH}).
This indicates that we cannot reliably obtain the mass of a cloud core without the effect of the magnetic field, because the larger density generally leads to smaller Jeans mass.

\begin{figure}[t]
\epsscale{1.}
\plotone{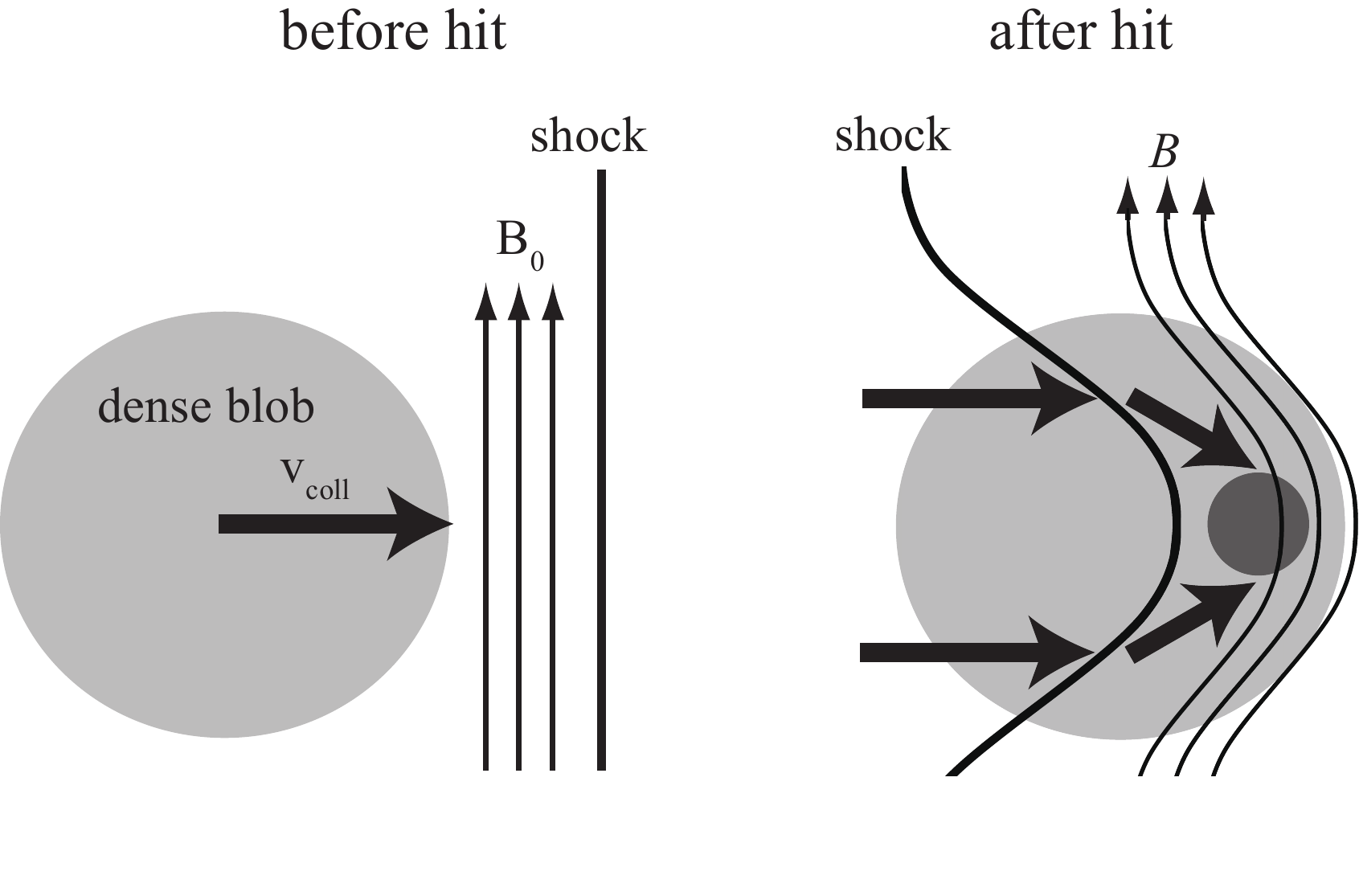}
\caption{\label{f1}
Schematics of the gas stream before ({\it left}) and after ({\it right}) the interaction between a shock and a dense blob.
Because the deformed shock wave kinks the stream lines across the shock, the stream lines are headed toward the convex point of the deformed shock wave.
}\end{figure}

\begin{figure}[t]
\epsscale{1.1}
\plotone{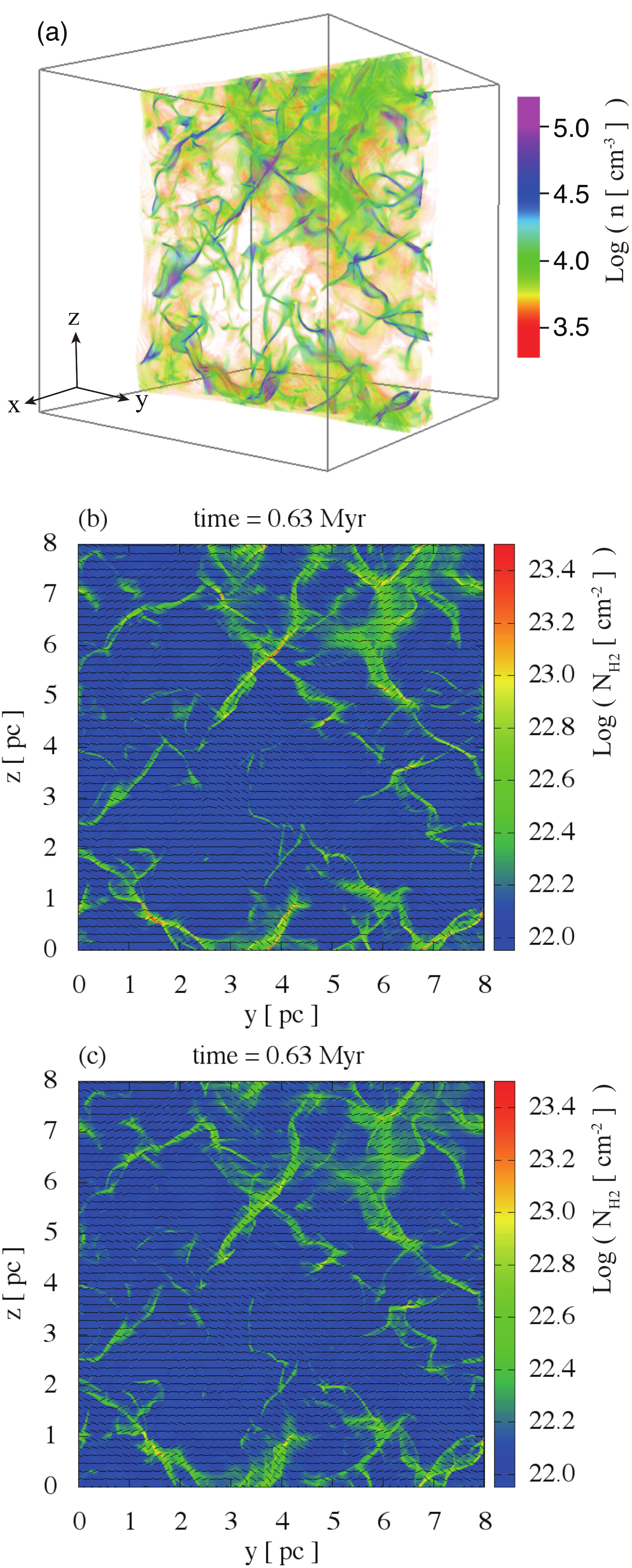}
\caption{\label{f2}
(a): volume rendering map of the density for Model 1 at $t=0.63$ Myr, in which the center of gravity of the massive core is located at $(x,\,y,\,z)=(4.0,\,6.1,\,7.1)$ pc.
(b): column density structure of the shocked gas integrated along the $x$-axis for Model 1 at $t=0.63$ Myr since the impact.
(c): same as panel (b) but for Model 3 (w/o gravity) at $t=0.63$ Myr.
The black bars in (b) and (c) indicate the orientation of the density-weighted mean magnetic field along the $x$-axis.
The ranges of the color bars in (b) and (c) are set from 75\% of the mean column density to 23.5 cm$^{-2}$.
The integration along the $x$-axis for the column density and the magnetic field orientation is performed only in the shocked layer.
}\end{figure}

In the present simulations, the upstream colliding clouds are inhomogeneous.
When the shock wave hits a dense blob in the cloud, the shock surface deforms owing to the enhanced ram pressure.
In Figure~\ref{f1}, we show the schematic of the gas stream when the fast shock is deformed\footnote{There are many works about the role of the interaction between a shock wave and the upstream density fluctuations in the astrophysics, e.g., Kornreich \& Scalo (2000); Gicalone \& Jokipii (2007); Kevlahan \& Pudritz (2009); Inoue et al. (2009, 2012); Inoue \& Inutuska (2012)}: 
Because the deformed shock wave kinks the stream lines across the shock, the stream lines are headed toward the convex point of the deformed shock wave (see also Vaidya et al 2013).
Owing to the effect of magnetic pressure, only the component of the kinked flows parallel to the postshock magnetic field lines can be focused and compressed.
This leads to the formation of a dense filament.
Figure~\ref{f2} shows the volume rendering map of the density (panel [a]) and column density structure of the shocked gas integrated along the $x$-axis (panel [b] and [c]) at $t=0.63$ Myr since the impact.
Note that the integrations along the $x$-axis for the column density and mean magnetic field are performed only in the shocked layer, which is defined as the region with $\rho > {\rm max}[\rho(t=0)]$.
As expected, we can see the formation of dense filaments in the shocked layer even in simulation without self-gravity (see panel [c]).
The black lines in panels (b) and (c) show the orientation of the density-weighted magnetic field projected onto the $y$-$z$ plane, indicating that the dense filaments are magnetized perpendicular to the filaments as expected.
Because the colliding two clouds have different orientation of magnetic fields normal to each other, which is the most probable case for random cloud-cloud collisions, the column density structure of the shocked gas shows cross-like features (i.e., filaments that are formed in the gas from $x>L_{\rm box}/2$ and $x<L_{\rm box}/2$ tend to be normal to each other).
Recent high-resolution observation of the massive star formation region in W33A showed that stars are formed at the intersection of two filaments (Galv\'{a}n-Madrid et al. 2010).
The result of our simulation reproduces such a characteristic site of massive star formation.

In Model 3 (w/o self-gravity), the postshock density never exceeds the value of $n_{\rm diag}\equiv 2\,\langle n\rangle_0 (v_{\rm coll}/c_{\rm s})^2$, because the thermal pressure of the gas with $n_{\rm diag}$ is twice the mean ram pressure of the colliding clouds.
In other words, it is diagnostic of the formation of a gravitationally bound object (core), if the density exceeds $n_{\rm diag}$.
To estimate the mass of the bound core, we use the following sequence by using data when the local density exceeds $n_{\rm diag}$:
First, identify the connected region with density larger than threshold density $n_{\rm thre}$ that involves the density peak of $n=n_{\rm diag}$.
Second, calculate the effective Jeans mass of the identified region
\begin{equation}
M_{\rm J, eff} \equiv \frac{\langle c_{\rm s} \rangle^3+\langle c_{\rm A} \rangle^3+\langle \Delta v \rangle^3}{G^{3/2}\,\langle \rho \rangle^{1/2}},
\end{equation}
where $\Delta v$ is the velocity dispersion.
Finally, if the mass of the region $M_{\rm core}$ is smaller than $M_{\rm J, eff}$, $n_{\rm thre}$ is increased until $M_{\rm core}$ becomes identical to $M_{\rm J, eff}$.

In the fiducial model, the massive core of $M_{\rm core}= 194 M_{\rm sun}$ is formed at $(x,y,z)=(4.0,\,6.1,\,7.1)$ pc and $t=0.63$ Myr.
The average density of the core is $\langle n \rangle =8.4\times10^4$ cm$^{-3}$ indicating that the thermal Jeans mass of the core is $M_{\rm J, therm} = c_{\rm s}^3\,G^{-3/2}\,\langle \rho \rangle^{-1/2}= 0.40$ M$_{\rm sun}$.
The ratios of the Alfv\'enic and turbulent Jeans mass to the thermal Jeans mass are $M_{\rm J, Alf}/M_{\rm J, therm}=\langle c_{\rm A} \rangle^3/\langle c_{\rm s} \rangle^3=333$ and $M_{\rm J, turb}/M_{\rm J, therm}=\langle \Delta v \rangle^3/\langle c_{\rm s} \rangle^3=196$, respectively.
Thus, the effective Jeans mass of the massive core is mainly enhanced by the strong magnetic field in the shocked layer.
Note that in the fiducial model, the mean magnetic field strengths of the massive core and the shocked layer are $|\langle \vec{B} \rangle_{\rm core}|=280\,\mu$G and $|\langle \vec{B} \rangle_{\rm sh}|=178\,\mu$G, respectively.

\begin{figure}[t]
\epsscale{1.15}
\plotone{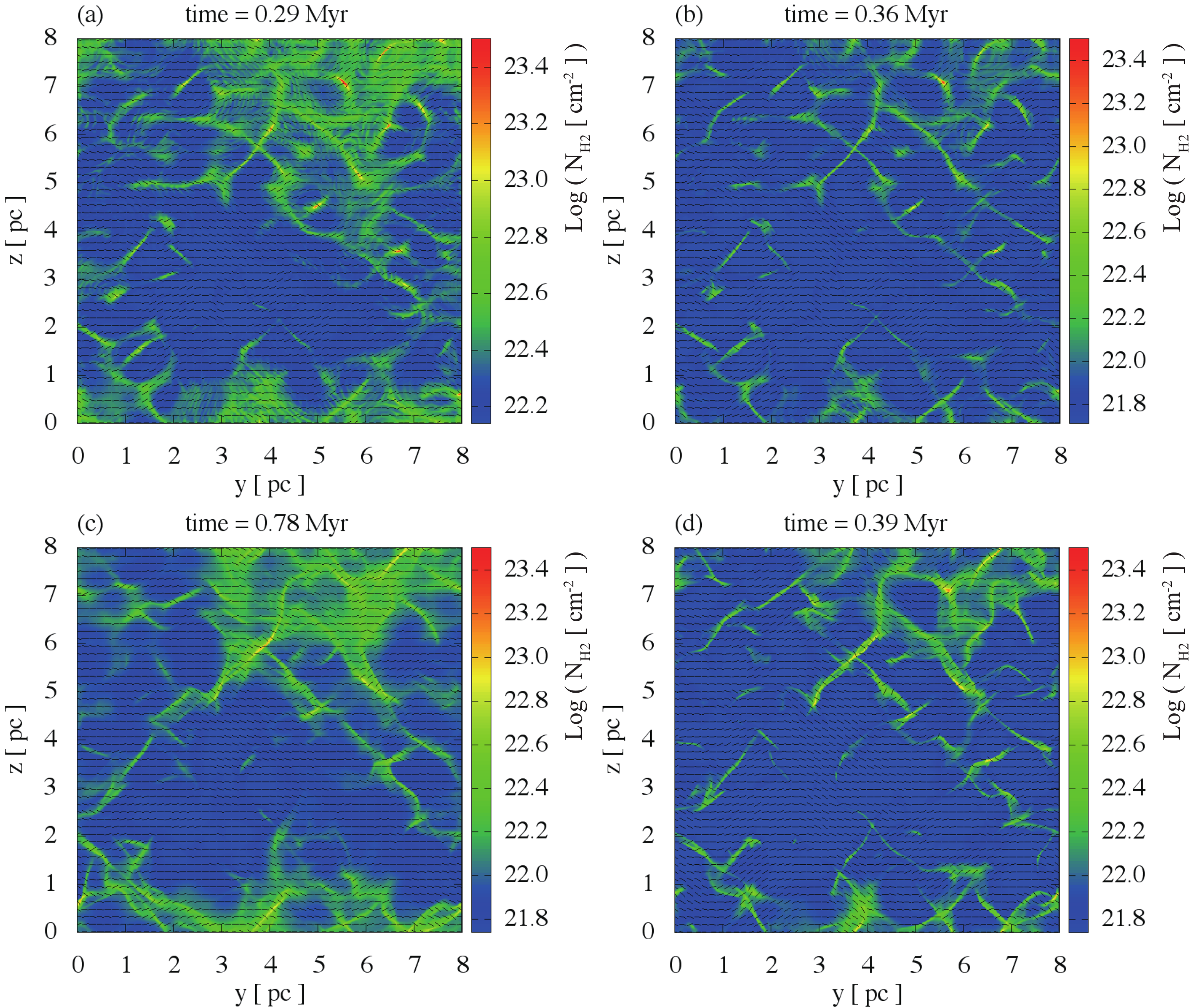}
\caption{\label{f3}
(a): column density structure of the shocked gas integrated along the $x$-axis for Model 4.
(b)-(d): same as (a) but for Models 5-7, respectively.
The ranges of the color bars are set from 75\% of the mean column density to 23.5 cm$^{-2}$.
The black bars indicate the orientation of the density-weighted mean magnetic field along the $x$-axis.
The integration along the $x$-axis for the column density and mean magnetic field is performed only in the shocked layer.
All panels show snapshots of the time when the massive core is formed at $(y,\,z)\sim(5.9,7.2)$ pc.
}\end{figure}

\subsection{Parameter dependency}
In Table 1, we summarize the physical parameters of the massive cores obtained by using the results of the other models.
The column density structures of Models 4-7 are shown in Figure~\ref{f3}.
Note that we focus only on the core formed at $(x,\,y,\,z)=(4.1\pm0.3,\,5.9\pm0.3,\,7.2\pm0.3)$, where the massive core is also formed in Model 1.
Thus, we can appropriately observe the effects of the initial parameters on the core mass.

The lower-resolution result (Model 2) shows that the mass of the massive core is apparently smaller than that of the fiducial model.
However, the difference is mostly attributed to the difference in the velocity dispersion (and consequently the turbulent Jeans mass) of the cores that is suppressed by numerical diffusion in the lower-resolution run.
The magnetic field strength and density of the core in Models 1 and 2 well coincide within 2\%.
This indicates that the Alfv\'enic Jeans mass is reliable even in the lower-resolution simulations, whereas the turbulent Jeans mass can be underestimated.
This result is consistent with the results of Heitsch et al. (2001) and Federrath et al. (2010), which showed that tens of numerical cells are necessary to resolve turbulent eddies.

The fiducial model result suggests that the mass of the massive core can be roughly estimated by the Alfv\'enic Jeans mass: $M_{\rm J, eff}\sim M_{\rm J, Alf}\propto \langle B\rangle_{\rm core}^3/\langle \rho\rangle_{\rm core}^2$.
If we substitute the mean density and magnetic field strength in the shocked layer given by Eq.~(\ref{RH}) in the above expression, we obtain $M_{\rm J, Alf}\propto B_0^2\,v_{\rm coll}\,\rho_0^{-3/2}$.
Because the core density is much larger than the mean density of the shocked layer due to the focused postshock flow (Figure~\ref{f1}), the above expression for the core mass is not accurate.
However, given that the mean magnetic field strength and mean density in the shocked layer correspond to the lower bounds of the core field strength and density, we can expect that the mass of the massive core is an increasing function of the initial magnetic field strength and the collision velocity, and a decreasing function of the initial density.
The results of Models 4 to 6 show that the mass of the massive cores is 126 M$_{\rm sun}$ for Model 4, 35 M$_{\rm sun}$ for Model 5, and 24 M$_{\rm sun}$ for Model 6 that are consistent with the above expectations.

The result of Model 7 (larger degree of initial density fluctuations) shows a smaller core mass than the fiducial model.
The reason for this is that the dense filament is formed by the postshock focusing flow (see, Figure~\ref{f1}), and the density of the filament becomes larger as the amplitude of the shock deformation increases.
Thus, the stronger interaction between the shock and density fluctuation in Model 7 leads to a higher filament density that results in a smaller core mass than the fiducial model.

The column density structures of Models 4-7 are shown in Figure~\ref{f3}.
Interestingly enough, despite the different column density, the thickness of the dense filaments are surprisingly uniform with $\simeq$ 0.1 pc irrespective of the initial parameters.
This invokes the recent discovery of 0.1 pc filaments in molecular clouds by Herschel space telescope (Arzoumanian et al. 2011).
More detailed analysis about the distribution of the thickness will be presented in our subsequent paper.

\section{Summary and Discussion}
By using three-dimensional isothermal MHD simulations, we have studied the formation of the massive molecular cloud core in the shocked layer induced by the intensive cloud-cloud collision.
Our findings are as follows:
\begin{itemize}
\item The collision of inhomogeneous clouds generates dense filaments in the shocked layer, and the direction of the magnetic field is perpendicular to the filaments.
The direction of the magnetic field is perpendicular to the filaments, and the thickness of the filaments apparently takes uniform value of $\simeq 0.1$ pc irrespective of the initial parameters.
\item The effective Jeans mass of the massive filament (core) is mostly determined by the Alfv\'enic and turbulent Jeans mass that is orders of magnitude larger than the thermal Jeans mass.
\item The mass of the massive core is an increasing function of the initial magnetic field strength and collision velocity and a decreasing function of the initial density.
These tendencies can be simply understood from the shock jump conditions and the Alfv\'enic Jeans mass.
\end{itemize}
The formation of the filaments indicates that the dense blob in the colliding cloud is multi-dimensionally compressed except perpendicular to the magnetic field by a single shock passage, because it is similar to the implosion process in a strongly magnetized medium.
Such multi dimensional compression enable it to generate a high column density and high mass object.

In the present simulations, we stopped the integration when the gravitationally bound massive core was formed.
We cannot conclude with certainty that massive stars are formed by the gravitational collapse of massive cores, because of the radiation feedback (Wolfire \& Cassinelli 1987) and the fragmentation by turbulence (Dobbs et al. 2005).
Several theoretical studies showed that massive stars can be formed by the gravitational collapse of a massive core, if the column density of the core is larger than $N_{{\rm H}_2}\gtrsim 1$ g cm$^{-2}$ $\simeq 10^{23.4}$ cm$^{-2}$ owing to radiative heating (Krumholz \& McKee 2008) and/or if the massive core is magnetized strongly (Commercon et al. 2011; Myers et al. 2013). 
However, Peters et al. (2011) and Hennebelle et al. (2011) showed that the level of fragmentation typically decreases only by a factor of 2.
As discussed by Girichidis et al. (2011), whether a massive core successfully produces massive star(s) would depend on the detailed structure of the core.
Peretto et al. (2007) showed by using SPH simulations that the gravitational collapse of an elongated core (or filament) could account for the massive cluster-forming clump NGC 2264-C.
The massive filamentary cores created in our simulations can be used as a realistic initial condition of massive star formation.

In this study, we have focused only on massive core formation.
However, the formation of dense filaments in the shocked layer can also be important in low-mass star formation.
Recent MHD simulations that consider detailed atomic/molecular cooling and ultraviolet heating with appropriate shielding effects showed that molecular clouds are intrinsically clumpy, and clump-clump collision is the basic process for forming gravitationally bound objects (Inoue \& Inutsuka 2012).
Thus, similar formation process of dense filaments is expected by the clump-clump collisions in a molecular cloud, suggesting that the 0.1 pc filaments discovered by Herschel telescope (Andr\'e et al. 2010; Arzoumanian et al. 2011) are possibly generated by the mechanism found in this paper.

\acknowledgments
We thank S. Inutsuka, K. Omukai, and T. Hosokawa for helpful discussions.
Numerical computations were carried out on XT4 and XC30 system at the Center for Computational Astrophysics (CfCA) of National Astronomical Observatory of Japan and K computer at the RIKEN Advanced Institute for Computational Science (No. hp120087).
This work is supported by Grant-in-aids from the Ministry of Education, Culture, Sports, Science, and Technology (MEXT) of Japan, No. 23740154.

\end{document}